\documentclass[11pt,a4paper]{article}
\usepackage[utf8]{inputenc}
\usepackage[T1]{fontenc}
\usepackage{lmodern}
\usepackage{amsmath,amssymb}
\usepackage{graphicx,booktabs}
\usepackage[margin=2.5cm]{geometry}
\usepackage{xcolor}
\definecolor{linkblue}{RGB}{25,70,140}
\usepackage[colorlinks=true,linkcolor=linkblue,citecolor=linkblue]{hyperref}
\usepackage{caption}
\newcommand{\sg}{\sqrt{-g}}
\newcommand{\dd}{\mathrm{d}}
\newcommand{\p}{\phi^{0}}

\title{\textbf{Cosmological Evolution of the Higgs Vacuum\\ Expectation
Value in Emergent Gravity}}
\author{Naz{\i}m R\"uzgar G\"uneri\\
\small Bah\c{c}e\c{s}ehir College, Antalya Parkorman Campus\\
\small\texttt{nazimruzagarguneri@gmail.com}
\and Metin Ar{\i}k\\
\small \.Istinye University\\
\small Bo\u{g}azi\c{c}i University\\
\small\texttt{metin.arik@istinye.edu.tr}\\
\small\texttt{metin.arik@boun.edu.tr}}
\date{\small August 24, 2026}

\begin{document}
\maketitle

\begin{abstract}
In a generally covariant theory whose Lagrangian density contains no
derivatives of the metric, the spacetime geometry is fixed algebraically by
the scalar fields, and the simplest quadratic potential is known to generate
exponentially expanding, recollapsing and big-rip cosmologies. We extend this
emergent-gravity Higgs cosmology by one new spacelike scalar field --- the
vacuum-expectation-value generating field --- coupled to the Higgs particle
through the Standard Model potential, so that the electroweak vacuum
expectation value becomes a dynamical quantity evolving with cosmological
time. We show that the broken-symmetry valley of the potential is an exact
and stable solution of the full nonlinear system, on which the background is
solved in closed form in terms of a single conserved charge. The charge feeds
the Friedmann equation with a radiation-like component of purely geometric
origin, diluting with the inverse fourth power of the scale factor although no
radiation fluid is present; as a consequence the universe is born at a finite
cosmological time in a radiation-dominated big bang, and for vanishing
constant term in the potential the expansion history realizes a spontaneous
radiation-to-dark-energy sequence. The sign of the Higgs kinetic term is not
an assumption of the model: the requirement that the broken phase be the
physical vacuum fixes it dynamically, and the stable choice reproduces the
Standard Model relation between the Higgs mass, the self-coupling and the
vacuum expectation value with no adjustable parameter. Oscillations about the
valley describe a Higgs condensate which is frozen at early times, begins to
oscillate through a misalignment mechanism, obeys a parameter-free adiabatic
amplitude law, and backreacts on the expansion history as a definite
renormalization of the background constants that feeds the radiation-like
component. All results are verified independently by computer algebra and by
high-precision numerical integration.
\end{abstract}

\section{Introduction}\label{sec:intro}

The idea that the gravitational field may not be elementary --- that the
metric of spacetime could be composed of, or induced by, matter --- goes back
to Sakharov's induced gravity \cite{Sakharov1967} and the pregeometry
programme of Akama \cite{Akama1978,AkamaHattori2013}, and reappears in the
induced-gravity symmetry-breaking approaches of Zee and Adler
\cite{Zee1979,Adler1982}; see \cite{Visser2002} for a modern perspective and
\cite{Wetterich2022} for a current pregeometry programme. Its sharpest early
incarnation is the proposal of Amati and Veneziano that the metric itself be
built from derivatives of matter fields \cite{AmatiVeneziano1981}. A parallel
line, initiated by Regge and Teitelboim \cite{ReggeTeitelboim1975} and
embodied in the Polyakov description of extended objects
\cite{Polyakov1981,DeserZumino1976}, treats spacetime as a surface embedded in
a higher-dimensional flat space
\cite{RubakovShaposhnikov1983,OverduinWesson1997,ErgenArik2024}. Emergent
metrics built from scalar gradients are, moreover, not merely a theoretical
speculation: the acoustic geometries of analogue gravity are precisely of this
type and are realized in the laboratory \cite{Unruh1981}. In recent years
this circle of ideas has been developed into a quantitative quantum theory:
in models where scalar fields serve as clocks and rulers and the metric is
determined by the constraint of vanishing total energy--momentum, the
quantized theory generates long-range gravitational interactions consistent
with general relativity, including graviton self-interactions
\cite{CaroneErlichVaman2017,ChaurasiaErlichZhou2018,CaroneClaringboldVaman2018},
and a stochastic framework exists into which any Lorentz-invariant field
theory, including the Standard Model, can be embedded \cite{Erlich2018}.

Following \cite{AyferArik2025,ErgenArik2024} we consider the extreme
classical version of these ideas: a generally covariant action that contains
\emph{no} derivatives of the metric whatsoever. The metric is then an
auxiliary field; its equation of motion is algebraic and can be solved once
and for all,
\begin{equation}
g_{\mu\nu}=\frac{f_{ab}\,\partial_\mu\phi^a\partial_\nu\phi^b}{V(\phi)},
\nonumber
\end{equation}
so that spacetime geometry is literally created by scalar fields. Related
constructions in which a metric or the cosmological dynamics is built from
scalars include disformal geometry \cite{Bekenstein1993}, mimetic gravity
\cite{Mimetic2013}, the ghost condensate \cite{Ghost2004}, k-essence
\cite{Kessence2000}, quintessence
\cite{RatraPeebles1988,Wetterich1988,CDS1998,Copeland2006}, modified-measure
theories \cite{GuendelmanKaganovich1999}, and scalar fields used as dynamical
standards of space and time \cite{BrownKuchar1995}. The present framework
differs from all of these in that the metric carries no kinetic term at all:
no Einstein--Hilbert term is present, and the metric equation is equivalent
to the statement that the \emph{total} energy--momentum tensor of the
universe vanishes \cite{HartleHawking1983,CaroneErlichVaman2017,Erlich2018}.
In this picture the time-creating scalar plays the role that the Higgs field
plays for particle masses, and the simplest quadratic potential already
produces exponentially expanding cosmologies as well as recollapsing and
big-rip ones \cite{AyferArik2025}.

The purpose of this paper is to introduce one new ingredient into the model
of \cite{AyferArik2025} and to work out its consequences completely. We
promote the electroweak vacuum expectation value to a dynamical field --- the
\emph{VEV-generating field} $\rho$ --- by coupling the Higgs-particle
component $h$ through the Standard Model potential
$\lambda(h^2-\rho^2/2)^2$, which is the usual $\lambda(|\Theta|^2-v^2/2)^2$
with the constant $v$ replaced by $\rho(t)$. A cosmological evolution of the
Higgs vacuum expectation value has been studied within Einstein gravity by
Calmet and collaborators
\cite{Calmet2017,CalmetFritzsch2002,CalmetKeller2015}, where it arises as an
effect on a fixed gravitational background; here, by contrast, the vacuum
expectation value is carried by a field that participates in creating the
geometry itself. The idea that cosmological evolution determines the
electroweak scale also has a celebrated mainstream counterpart in the
relaxion \cite{GKR2015}; the mechanism here is different --- the vacuum
tracks the field exactly instead of being scanned --- and the gravitational
side is emergent rather than Einsteinian.

Our main results are the following. (i)~The broken-symmetry valley
$h^2=\rho^2/2$ is an exact, stable solution of the full nonlinear system,
and on it the background is solved in closed form in terms of a single
conserved charge. (ii)~This charge feeds the Friedmann equation with a
radiation-like $a^{-4}$ component of purely geometric origin --- no radiation
fluid is introduced --- in close analogy with the ``mirage'' radiation
produced by conserved brane motion in mirage cosmology
\cite{MirageCosmology1999}; as a consequence, for $C\neq0$ the universe
acquires a radiation-dominated birth at finite cosmological time.
(iii)~The sign of the Higgs kinetic term is not an assumption: the
requirement that the broken phase be the physical vacuum fixes it
dynamically, and the stable choice reproduces the Standard Model Higgs mass
relation $m_H^2=2\lambda v^2$ with no adjustable parameter. (iv)~The Higgs
condensate --- oscillations about the valley --- realizes a misalignment
mechanism \cite{PWW1983,Turner1983,Marsh2016} and backreacts on the
expansion as a definite renormalization of the background constants, feeding
the radiation component. The $V_0=0$ member of the solution family realizes
a spontaneous radiation~$\to$~dark-energy sequence, which is of natural
interest for cosmic acceleration
\cite{Riess1998,Perlmutter1999,PeeblesRatra2003} and for the current
discussions of evolving dark energy \cite{DESI2024,DESI2025} and of the
Hubble tension
\cite{Planck2018,Riess2022,Verde2019,DiValentino2021,CosmoVerse2025}.

The paper is organized as follows. Section~\ref{sec:framework} derives the
metric equation and the scalar field equations from the action.
Section~\ref{sec:model} defines the model and its field content.
Section~\ref{sec:gauge} fixes the gauge so that the time coordinate is the
cosmological time from the outset, reduces the field equations, and proves
the consistency identity that renders four of them dependent.
Section~\ref{sec:valley} establishes the valley as an exact solution of the
full nonlinear system and reduces the background to a single canonical
field. Section~\ref{sec:friedmann} derives the conserved charge, the
closed-form background, and the modified Friedmann equation, together with
its physical interpretation and asymptotics. Section~\ref{sec:sign}
determines the sign of the Higgs kinetic term and derives the Higgs mass
relation. Section~\ref{sec:condensate} studies the Higgs condensate and its
backreaction. Section~\ref{sec:verification} summarizes the verification
suite, and Section~\ref{sec:conclusions} collects the conclusions and open
problems. Appendix~\ref{app:static} presents the same background in the
static gauge, where the time-creating field is pinned to a timelike
coordinate $T$, and demonstrates the equivalence of the two descriptions;
Appendix~\ref{app:verification} itemizes the verification;
Appendix~\ref{app:ridge} treats the unstable symmetric solution.

Throughout we follow the notation and the dimensional conventions of
\cite{AyferArik2025}: spacetime coordinates $x^\mu$ have dimension of
inverse mass, all scalar fields have dimension of mass, and $t$ denotes the
cosmological time.

\section{The framework: gravity without metric derivatives}
\label{sec:framework}

\subsection{Action and the metric equation}

Following \cite{AyferArik2025}, we consider scalar fields $\phi^a$ living in
a flat configuration space with constant metric $f_{ab}$, coupled to a
spacetime metric $g_{\mu\nu}$ through the generally covariant action
\begin{equation}\label{eq:action}
\mathcal S[\phi^a,g_{\mu\nu}]
=\int \sg\,\Bigl[\tfrac12\,g^{\mu\nu} f_{ab}\,
\partial_\mu\phi^a\partial_\nu\phi^b
-V(\phi)\Bigr]\dd^4x ,
\end{equation}
where $V(\phi)$ is the field-dependent cosmological term. No curvature term
is present, so \eqref{eq:action} contains no derivatives of $g_{\mu\nu}$,
and the variation with respect to the metric is purely algebraic. Using
\begin{equation}\label{eq:deltasg}
\delta\sg=-\tfrac12\,\sg\,g_{\mu\nu}\,\delta g^{\mu\nu},
\end{equation}
the variation of \eqref{eq:action} with respect to $g^{\mu\nu}$ gives
\begin{equation}\label{eq:variation}
\frac{\delta\mathcal S}{\delta g^{\mu\nu}}
=\sg\Bigl[\tfrac12 f_{ab}\,\partial_\mu\phi^a\partial_\nu\phi^b
-\tfrac12 g_{\mu\nu}\Bigl(\tfrac12 g^{\alpha\beta}f_{ab}
\partial_\alpha\phi^a\partial_\beta\phi^b-V\Bigr)\Bigr]=0 .
\end{equation}
Taking the trace with $g^{\mu\nu}$ (four spacetime dimensions,
$g^{\mu\nu}g_{\mu\nu}=4$),
\begin{equation}\label{eq:trace}
\tfrac12\,g^{\mu\nu}f_{ab}\partial_\mu\phi^a\partial_\nu\phi^b
-2\Bigl(\tfrac12 g^{\alpha\beta}f_{ab}
\partial_\alpha\phi^a\partial_\beta\phi^b-V\Bigr)=0
\qquad\Longrightarrow\qquad
\tfrac12\,g^{\alpha\beta}f_{ab}
\partial_\alpha\phi^a\partial_\beta\phi^b=2V ,
\end{equation}
and substituting \eqref{eq:trace} back into \eqref{eq:variation} solves the
metric equation once and for all:
\begin{equation}\label{eq:metric}
\boxed{\;g_{\mu\nu}
=\frac{f_{ab}\,\partial_\mu\phi^a\,\partial_\nu\phi^b}{V(\phi)}\;.}
\end{equation}
This is Eq.~(2) of \cite{AyferArik2025} and the key equation of the
framework: the metric of spacetime is expressed in terms of the fields of
the configuration space, $\dd s^2_x=\dd s^2_\phi/V$. Note that the bracket
in \eqref{eq:variation} is one half of the total energy--momentum tensor of
the scalar fields,
\begin{equation}\label{eq:Tdef}
T_{\mu\nu}= f_{ab}\,\partial_\mu\phi^a\partial_\nu\phi^b
-g_{\mu\nu}\Bigl(\tfrac12\,g^{\alpha\beta}f_{ab}\,
\partial_\alpha\phi^a\partial_\beta\phi^b-V\Bigr),
\end{equation}
so the metric equation \eqref{eq:metric} is precisely the statement
$T_{\mu\nu}=0$: it can be said to guarantee zero total energy--momentum of
the universe \cite{HartleHawking1983}. The same constraint is the starting point of the
emergent-gravity models of
\cite{CaroneErlichVaman2017,ChaurasiaErlichZhou2018,%
CaroneClaringboldVaman2018,Erlich2018}, where its quantization was shown to
generate long-range gravity consistent with general relativity; here we
remain at the classical level and use it as an exact algebraic solution for
the metric.

\subsection{Scalar field equations}

Varying \eqref{eq:action} with respect to $\phi^a$ gives
\begin{equation}\label{eq:scalarEOM}
\partial_\mu\Bigl[\sg\, f_{ab}\,g^{\mu\nu}\partial_\nu\phi^b\Bigr]
+\sg\,\frac{\partial V}{\partial\phi^a}=0 ,
\end{equation}
which is Eq.~(3) of \cite{AyferArik2025}. Equations \eqref{eq:metric} and
\eqref{eq:scalarEOM} together define the dynamics: the metric is eliminated
algebraically through \eqref{eq:metric}, after which \eqref{eq:scalarEOM}
becomes a closed system for the scalar fields alone.

\section{The model}\label{sec:model}

\subsection{Field content}

For $g_{\mu\nu}$ to have Lorentzian signature $(+,-,-,-)$ with $V>0$, the
configuration space must be at least four dimensional, with $f_{ab}$
possessing at least one positive and at least three negative eigenvalues
\cite{AyferArik2025,Akama1978}. The minimal choice of \cite{AyferArik2025}
consists of a time-creating field $\p$ and three space-creating fields
$\phi^i$. To this we add the two fields of the electroweak sector: the
Higgs-particle component $h$ and the VEV-generating field $\rho$. The
configuration space is therefore six dimensional,
\begin{equation}\label{eq:fields}
\phi^a=\bigl(\p,\;h,\;\phi^1,\phi^2,\phi^3,\;\rho\bigr),
\qquad
f_{ab}=\mathrm{diag}\bigl(+1,\;s,\;-1,-1,-1,\;-1\bigr),
\qquad s=\pm1 ,
\end{equation}
and the Lagrangian density of \eqref{eq:action} reads
\begin{equation}\label{eq:lagrangian}
\mathcal L=\sg\,\Bigl[\tfrac12 g^{\mu\nu}\Bigl(
\partial_\mu\p\partial_\nu\p
+s\,\partial_\mu h\,\partial_\nu h
-\partial_\mu\phi^i\partial_\nu\phi^i
-\partial_\mu\rho\,\partial_\nu\rho\Bigr)-V\Bigr] ,
\end{equation}
with the potential
\begin{equation}\label{eq:potential}
\boxed{\;V=V_0+\tfrac12\,m^2(\p)^2
+\lambda\Bigl(h^2-\tfrac12\,\rho^2\Bigr)^{2}\;.}
\end{equation}

Three remarks on the structure of
\eqref{eq:fields}--\eqref{eq:potential} are in order.

(i) The term $V_0+\tfrac12 m^2(\p)^2$ is the quadratic potential of the
time-creating field, Eq.~(11) of \cite{AyferArik2025}, which by itself
generates the three cosmologies of that paper (exponential expansion for
$V_0=0$, expansion followed by collapse for $V_0>0$, big rip for $V_0<0$).
When the electroweak pair $(h,\rho)$ is switched off, the present model
reduces exactly to \cite{AyferArik2025}.

(ii) The $\lambda$ term is the Standard Model Higgs potential
$\lambda(|\Theta|^2-v^2/2)^2$ with the constant vacuum expectation value
replaced by the dynamical field, $v\to\rho(t)$. For other cosmological roles
of the Higgs field see Higgs inflation \cite{Bezrukov2008} and
Higgs--dilaton cosmology \cite{HiggsDilaton}, where a dynamical electroweak
scale also appears.

(iii) The fields $\phi^i$ and $\rho$ carry negative kinetic signature ---
phantom-type in Minkowski language \cite{Caldwell2002,CHT2003}. As
emphasized in \cite{AyferArik2025}, in this framework such \emph{spacelike
scalar fields} are not pathological but necessary: they generate the space
dimensions, and the vanishing of the total energy--momentum tensor removes
the usual instability argument. The sign $s$ of the $h$ kinetic term, by
contrast, is \emph{not} fixed by the signature counting; we shall show in
Section~\ref{sec:sign} that it is fixed by the dynamics.

\section{Gauge fixing, cosmological time and reduced equations}
\label{sec:gauge}

\subsection{Coordinates adapted to cosmology}

Since the action is generally covariant, four arbitrary coordinate choices
are allowed, provided the relevant Jacobian is nonzero \cite{AyferArik2025}.
Exactly as in Eq.~(7) of \cite{AyferArik2025}, we spend three of them by
pinning the space-creating fields to the spatial coordinates, and we use the
fourth to make the time coordinate the \emph{cosmological time}:
\begin{equation}\label{eq:gauge}
\phi^i=\mu^2 x^i ,
\qquad
\p=\p(t),\quad h=h(t),\quad\rho=\rho(t)
\quad\text{with $t$ such that}\quad g_{00}=1 ,
\end{equation}
where $\mu$ has dimension of mass. (Homogeneity of $\p$, $h$, $\rho$ is the
cosmological ansatz; the gauge content of \eqref{eq:gauge} is the choice of
$x^i$ and of $t$.) With \eqref{eq:gauge} the metric \eqref{eq:metric} is
diagonal, and its components read, with dots denoting $\dd/\dd t$,
\begin{equation}\label{eq:metricgauge}
g_{00}=\frac{\dot{\p}^{\,2}+s\,\dot h^{2}-\dot\rho^{\,2}}{V}=1,
\qquad
g_{0i}=0,
\qquad
g_{ij}=-\frac{\mu^4}{V}\,\delta_{ij}\equiv-a^2(t)\,\delta_{ij},
\end{equation}
so that the line element is the flat FLRW form
$\dd s^2=\dd t^2-a^2(t)\,\dd\vec r^{\,2}$ and the scale factor is
\begin{equation}\label{eq:scalefactor}
a(t)=\frac{\mu^2}{\sqrt{V}} ,
\end{equation}
exactly Eq.~(10) of \cite{AyferArik2025}: \emph{the relation between the
scale factor and the potential is unchanged by the new fields}. The first
equality in \eqref{eq:metricgauge} is the definition of cosmological time
and will be referred to as the \textbf{normalization condition},
\begin{equation}\label{eq:normalization}
\dot{\p}^{\,2}+s\,\dot h^{2}-\dot\rho^{\,2}=V .
\end{equation}
For the pure time-creating field ($\dot h=\dot\rho=0$) it reduces to
$\dd\p/\sqrt V=\dd t$, Eq.~(9) of \cite{AyferArik2025}. Determinant and
inverse metric are
\begin{equation}\label{eq:line}
\sg=a^3=\frac{\mu^6}{V^{3/2}},\qquad
g^{00}=1,\qquad g^{ij}=-\frac{V}{\mu^4}\,\delta^{ij} .
\end{equation}
Lorentzian signature requires $V>0$ and reality of $t$ in
\eqref{eq:normalization}; the solution constructed below preserves both
automatically. An alternative, equivalent gauge --- pinning $\p=\mu^2T$ to a
timelike coordinate $T$ and leaving $g_{00}\neq1$ --- is presented in
Appendix~\ref{app:static}; there $T$ is merely a coordinate label, not a
clock, and the conversion $\dd t=\sqrt{g_{00}}\,\dd T$ reproduces every
result of the main text. Scalar fields serving in this way as clocks and
rulers, and the associated distinction between coordinate labels and
physical time, are recurring themes of relational dynamics
\cite{PageWootters1983,BrownKuchar1995,Torre1992,Erlich2018}.

\subsection{Reduced field equations}

Substituting \eqref{eq:line} into \eqref{eq:scalarEOM}, each homogeneous
field obeys
$\dd\bigl[a^3 f_{aa}\dot\phi^a\bigr]/\dd t+a^3\,\partial V/\partial\phi^a=0$
(no sum), while the space-creating fields drop out. Field by field:
\begin{align}
\phi^i:&\quad
\partial_i\Bigl[\sg\,g^{ii}\mu^2\Bigr]=0
\qquad\text{identically, since $\sg\,g^{ii}$ depends only on $t$;}
\label{eq:Xred}\\[3pt]
\p:&\quad
\frac{\dd}{\dd t}\Bigl[a^3\,\dot{\p}\Bigr]
=-\,a^3\,m^2\,\p ,
\label{eq:phired}\\[3pt]
h:&\quad
s\,\frac{\dd}{\dd t}\Bigl[a^3\,\dot h\Bigr]
=-\,4\lambda\,a^3\,h\Bigl(h^2-\tfrac12\rho^2\Bigr),
\label{eq:hred}\\[3pt]
\rho:&\quad
\frac{\dd}{\dd t}\Bigl[a^3\,\dot\rho\Bigr]
=-\,2\lambda\,a^3\,\rho\Bigl(h^2-\tfrac12\rho^2\Bigr),
\label{eq:rhored}
\end{align}
where we used
$\partial V/\partial h=4\lambda h(h^2-\rho^2/2)$ and
$\partial V/\partial\rho=-2\lambda\rho(h^2-\rho^2/2)$; the negative kinetic
signature of $\rho$ has flipped the sign of its equation. These are the
standard homogeneous-scalar equations in an FLRW background --- e.g.\
\eqref{eq:hred} is $s(\ddot h+3H\dot h)=-\partial V/\partial h$ with
$H\equiv\dot a/a$ --- except that here $a=\mu^2/\sqrt V$ is itself built
from the fields, and the normalization condition \eqref{eq:normalization}
must hold as well.

\subsection{The consistency identity}\label{sec:identity}

The system
\eqref{eq:normalization}--\eqref{eq:rhored} appears overdetermined: four
differential equations and one constraint for three functions. It is not.
By Noether's second theorem, each local symmetry function generates one
identity among the equations of motion; having spent four coordinate
functions, exactly four equations must become dependent. The three $\phi^i$
equations are identically satisfied, and the $\p$ equation
\eqref{eq:phired} is automatically satisfied on the shell of the remaining
equations. On the valley solution of Section~\ref{sec:valley} this can be
displayed in one line (we do so in Section~\ref{sec:friedmann}); in general
it follows from the diffeomorphism identity of the covariant system, which
we have verified symbolically for arbitrary field profiles
(Appendix~\ref{app:verification}). The counting closes exactly: six fields
minus four gauge functions leave \emph{two} physical degrees of freedom ---
the background mode and the Higgs oscillation studied below.

\section{Symmetry breaking: the valley as an exact solution}\label{sec:valley}

\subsection{Critical structure of the potential}

We first locate the extrema of the potential \eqref{eq:potential} in the $h$
direction at fixed $\rho$. Differentiating,
\begin{equation}\label{eq:Vh}
\frac{\partial V}{\partial h}
=4\lambda\,h\Bigl(h^2-\tfrac12\rho^2\Bigr)=0
\qquad\Longrightarrow\qquad
h=0\quad(\text{ridge})
\qquad\text{or}\qquad
h^2=\tfrac12\rho^2\quad(\text{valley}),
\end{equation}
with curvatures
\begin{equation}\label{eq:Vhh}
\frac{\partial^2V}{\partial h^2}
=4\lambda\Bigl(3h^2-\tfrac12\rho^2\Bigr)
=\begin{cases}
-2\lambda\rho^2<0 & \text{at }h=0 \quad(\text{maximum}),\\[4pt]
+4\lambda\rho^2>0 & \text{at }h^2=\tfrac12\rho^2 \quad(\text{minimum}).
\end{cases}
\end{equation}
These are the symmetric and broken phases of the Standard Model potential,
with the constant vacuum expectation value replaced by the VEV-generating
field: on the valley the physical, time-dependent electroweak scale is
\begin{equation}\label{eq:vev}
v(t)=\rho(t) .
\end{equation}

\subsection{Exactness of the valley embedding}

Because $h$ is dynamical, it must be verified that the embedding
\begin{equation}\label{eq:embedding}
h(t)=\frac{\varepsilon}{\sqrt2}\,\rho(t),
\qquad\varepsilon=\pm1 ,
\end{equation}
solves the coupled system exactly. On \eqref{eq:embedding} the combination
$h^2-\rho^2/2$ vanishes identically, so
\begin{equation}
\frac{\partial V}{\partial h}\bigg|_{\rm valley}
=\frac{\partial V}{\partial\rho}\bigg|_{\rm valley}=0 ,
\qquad
V\big|_{\rm valley}=V_0+\tfrac12\,m^2(\p)^2 ,
\end{equation}
i.e.\ the $\lambda$ term drops from the potential itself, and the two
equations \eqref{eq:hred} and \eqref{eq:rhored} reduce to
\begin{equation}\label{eq:tworeduced}
s\,\frac{\dd}{\dd t}\Bigl[a^3\,\dot h\Bigr]=0,
\qquad
\frac{\dd}{\dd t}\Bigl[a^3\,\dot\rho\Bigr]=0 .
\end{equation}
Since $\dot h=\varepsilon\dot\rho/\sqrt2$ on \eqref{eq:embedding}, the first
equation is $\varepsilon s/\sqrt2$ times the second: the two collapse
consistently into \emph{one}. The valley is thus an exact invariant
submanifold of the full nonlinear system --- confirmed both symbolically and
numerically: the integrated full system stays on the valley to
$3.7\times10^{-12}$ (Appendix~\ref{app:verification}). The ridge $h=0$ is
an exact solution by the same argument; being a maximum of $V$ it is
unstable (Section~\ref{sec:sign}) and is treated in
Appendix~\ref{app:ridge}.

\subsection{The canonical valley field and form invariance}

On the valley the kinetic combination of the pair $(h,\rho)$ collapses.
Substituting $\dot h=\varepsilon\dot\rho/\sqrt2$ with $s=+1$ (the physical
sign, as established in Section~\ref{sec:sign}),
\begin{equation}\label{eq:kineticalgebra}
\dot h^{2}-\dot\rho^{\,2}
=\tfrac12\,\dot\rho^{\,2}-\dot\rho^{\,2}
=-\tfrac12\,\dot\rho^{\,2}
=-\,\dot\sigma^{\,2},
\qquad
\boxed{\;\sigma\equiv\frac{\rho}{\sqrt2}=\varepsilon\,h\;,}
\end{equation}
so the normalization condition \eqref{eq:normalization} becomes
\begin{equation}\label{eq:normvalley}
\dot{\p}^{\,2}-\dot\sigma^{\,2}=V .
\end{equation}
Valley motion is therefore described by a \emph{single} canonically
normalized spacelike field $\sigma$. This is a statement of \textbf{form
invariance}: on the valley, the model \emph{with} the Higgs kinetic term is
mapped exactly onto the model \emph{without} it under $\rho\to\sigma$; every
background formula below holds in both models, with the dictionary
$v=\rho=\sqrt2\,\sigma$ relating the physical electroweak scale to the
canonical field.

\section{The background in closed form and the modified Friedmann equation}
\label{sec:friedmann}

\subsection{The conserved charge}

On the valley the surviving field equation \eqref{eq:tworeduced} is a total
derivative --- the Noether current of the shift symmetry
$\sigma\to\sigma+\mathrm{const}$ of the valley-reduced action. Its first
integral defines the conserved charge $C$ of the VEV-generating field,
\begin{equation}\label{eq:charge}
\boxed{\;a^3\,\dot\sigma=\mu^6\,C=\text{const}\;,}
\qquad [C]=M^{-4} .
\end{equation}
With $a^3=\mu^6/V^{3/2}$ from \eqref{eq:line}, the charge determines the
velocity of the VEV-generating field algebraically,
\begin{equation}\label{eq:sigmadot}
\dot\sigma=C\,V^{3/2},
\end{equation}
and the normalization condition \eqref{eq:normvalley} then fixes the
velocity of the time-creating field,
\begin{equation}\label{eq:phidot}
\dot{\p}^{\,2}=V+\dot\sigma^{\,2}=V+C^2V^3=V\bigl(1+C^2V^2\bigr) .
\end{equation}
Equations \eqref{eq:sigmadot}--\eqref{eq:phidot}, together with
$V=V_0+\tfrac12 m^2(\p)^2$, solve the background completely: given $C$, the
single function $\p(t)$ is determined by the first-order equation
\eqref{eq:phidot}, and $\sigma$, $a$, $H$ follow algebraically. Comparing
with Eq.~(9) of \cite{AyferArik2025}, $\dd\p/\sqrt V=\dd t$, the entire
effect of the VEV-generating field on the background is the factor
$\sqrt{1+C^2V^2}$ in \eqref{eq:phidot}.

\subsection{Consistency check: the time-creating field equation is
automatic}

Before extracting the cosmology we display, in one line, the consistency
identity promised in Section~\ref{sec:identity}. Differentiating the
normalization \eqref{eq:phidot} and using $\dot V=m^2\p\,\dot{\p}$,
\begin{equation}
2\,\dot{\p}\,\ddot{\p}
=\dot V\bigl(1+3C^2V^2\bigr)
\qquad\Longrightarrow\qquad
\ddot{\p}=\tfrac12\,m^2\,\p\,\bigl(1+3C^2V^2\bigr),
\end{equation}
so the left-hand side of the $\p$ equation \eqref{eq:phired} becomes
\begin{align}
\frac1{a^3}\frac{\dd}{\dd t}\Bigl[a^3\dot{\p}\Bigr]
&=\ddot{\p}-\frac{3\,\dot V}{2\,V}\,\dot{\p}
=\tfrac12 m^2\p\bigl(1+3C^2V^2\bigr)
-\frac{3\,m^2\p\,\dot{\p}^{\,2}}{2\,V}
\nonumber\\[2pt]
&=\tfrac12 m^2\p\Bigl[\bigl(1+3C^2V^2\bigr)
-3\bigl(1+C^2V^2\bigr)+2\Bigr]-m^2\p
=-\,m^2\,\p ,
\label{eq:identitycheck}
\end{align}
which is exactly the right-hand side: the $\p$ equation holds identically,
with no condition on $C$. (In the middle step
$\dot{\p}^{\,2}/V=1+C^2V^2$ from \eqref{eq:phidot} was used.) This is the
concrete form of the Noether identity in the cosmological gauge; we have
also verified it symbolically (Appendix~\ref{app:verification}).

\subsection{The modified Friedmann equation}

The Hubble parameter follows from \eqref{eq:scalefactor} by pure
differentiation. Every step is elementary but we display them all. From
$a=\mu^2V^{-1/2}$,
\begin{equation}
H\equiv\frac{\dot a}{a}=-\frac{\dot V}{2V}
=-\frac{m^2\,\p\,\dot{\p}}{2V},
\qquad
H^2=\frac{m^4\,(\p)^2\,\dot{\p}^{\,2}}{4V^2}
=\frac{m^4\,(\p)^2\,\bigl(1+C^2V^2\bigr)}{4V},
\end{equation}
using \eqref{eq:phidot} in the last step. Eliminating the field through
$m^2(\p)^2=2(V-V_0)$,
\begin{equation}\label{eq:H2V}
H^2=\frac{m^2}{2}\Bigl(1-\frac{V_0}{V}\Bigr)\bigl(1+C^2V^2\bigr)
=\alpha^2\Bigl(1-\frac{V_0}{V}\Bigr)\bigl(1+C^2V^2\bigr),
\qquad
\alpha\equiv\frac{m}{\sqrt2}\,,
\end{equation}
where $\alpha$ is the expansion constant of \cite{AyferArik2025}. Finally,
eliminating $V$ in favour of the observable scale factor through
$V=\mu^4/a^2$, and introducing the turnaround radius $a_0=\mu^2/\sqrt{V_0}$
(for $V_0>0$), we obtain the central result of this paper:
\begin{equation}\label{eq:friedmann}
\boxed{\;H^2=\alpha^{2}
\Bigl(1-\frac{a^2}{a_0^{2}}\Bigr)\Bigl(1+\frac{C^2\mu^8}{a^4}\Bigr)
=\alpha^{2}\biggl[\,1
+\underbrace{\frac{C^2\mu^8}{a^4}}_{\text{radiation-like}}
-\underbrace{\frac{C^2\mu^8}{a_0^2\,a^2}}_{\text{curvature-like}}
-\underbrace{\frac{a^2}{a_0^{2}}}_{\text{turnaround}}\,\biggr].\;}
\end{equation}
For $C=0$ this collapses to $H^2=\alpha^2(1-a^2/a_0^2)$, whose solutions
are precisely Table~1 of \cite{AyferArik2025}: $a=Ae^{\alpha t}$ for
$V_0=0$, $a=A/\cosh(\alpha t)$ for $V_0>0$, and $a=A/\sinh(\alpha t)$ (big
rip) for $V_0<0$.

\subsection{Physical reading of the four components}

Equation \eqref{eq:friedmann} has the additive structure of a standard
Friedmann budget, $H^2\propto R/a^4+S/a^2+\Lambda+\dots$, with
\begin{equation}
R\;\leftrightarrow\;\alpha^2C^2\mu^8,
\qquad
S\;\leftrightarrow\;-\,\alpha^2C^2\mu^8/a_0^2,
\qquad
\Lambda\;\leftrightarrow\;\alpha^2,
\end{equation}
plus the turnaround term $-\alpha^2a^2/a_0^2$ special to this model. Yet
\emph{no} radiation fluid, spatial curvature or cosmological constant has
been introduced: a single conserved charge of a homogeneous scalar generates
the $a^{-4}$ and $a^{-2}$ terms geometrically. Three contrasts sharpen the
point.

(i) A minimally coupled scalar in Einstein gravity whose energy is
kinetic-dominated redshifts as the \emph{stiff} $a^{-6}$ (kination
\cite{Spokoiny1993,Joyce1997}); the $a^{-4}$ found here is softer and is a
distinctive signature of the emergent-metric framework, traceable to the
$1/V$ structure of \eqref{eq:metric}.

(ii) The closest known analogue is \emph{mirage cosmology}
\cite{MirageCosmology1999}, where the conserved momentum of a brane moving
through a bulk produces a term in the induced Friedmann equation
indistinguishable from dilute radiation. The charge \eqref{eq:charge} plays
exactly the role of the conserved brane momentum, with the configuration
space of \eqref{eq:fields} as the bulk.

(iii) The $a^{-2}$ term enters with the sign of \emph{positive} spatial
curvature for $V_0>0$, although the spatial sections of
\eqref{eq:metricgauge} are exactly flat: an effective curvature term of
dynamical origin.

\subsection{Asymptotics: a radiation-dominated birth}

Whenever $C^2V^2\gg1$ --- that is, at small scale factor,
$a\ll a_\times\equiv\mu^2\sqrt C$ --- the radiation-like term dominates
\eqref{eq:friedmann}:
\begin{equation}
H^2\simeq\alpha^2\,\frac{C^2\mu^8}{a^4}
\qquad\Longrightarrow\qquad
a\,\frac{\dd a}{\dd t}\simeq\alpha\,C\mu^4
\qquad\Longrightarrow\qquad
a^2(t)\simeq2\,\alpha\,C\mu^4\,(t-t_b),
\label{eq:radiation}
\end{equation}
i.e.\ $a\propto(t-t_b)^{1/2}$, the exact radiation-era law. Its consequence
is
qualitative: for any $C\neq0$ the universe is born at a \emph{finite}
cosmological time $t_b$ in a radiation-dominated big bang ($a\to0$,
$H\to\infty$), whereas for $C=0$ the point $a=0$ is approached only as
$t\to-\infty$. The VEV-generating field converts an eternal past into a hot
birth.

For $V_0=0$ the crossover from radiation domination to dark-energy
domination occurs at $C^2V^2=1$, and the cosmological time elapsed between
the big bang and this crossover can be computed in closed form. From
\eqref{eq:H2V} with $V_0=0$, $\dot V=-2VH=-2\alpha V\sqrt{1+C^2V^2}$ on the
expanding branch, so with the substitution $u=CV$,
\begin{equation}\label{eq:age}
t_\times-t_b
=\int_{1/C}^{\infty}\frac{\dd V}{2\alpha V\sqrt{1+C^2V^2}}
=\frac1{2\alpha}\int_{1}^{\infty}\frac{\dd u}{u\sqrt{1+u^2}}
=\boxed{\;\frac{\ln\bigl(1+\sqrt2\bigr)}{\sqrt2\,m}\;}
\end{equation}
--- \emph{independent of the charge}: in this model the age of the universe
at radiation/dark-energy equality is a universal constant set by the mass of
the time-creating field alone. The subsequent history is classified by the
sign of $V_0$ in Table~\ref{tab:V0}; the $V_0=0$ member realizes, with no
additional ingredient, the sequence radiation era~$\to$~de Sitter
acceleration, and the $V_0<0$ member ends in a big rip
\cite{Caldwell2002,CKW2003,NOT2005}.

\begin{table}[t]
\centering
\small
\begin{tabular}{@{}lll@{}}
\toprule
& $C=0$ (Ayfer--Ar{\i}k \cite{AyferArik2025}) & $C\neq0$ (with VEV-generating field) \\
\midrule
$V_0>0$ & $a=A/\cosh(\alpha t)$, infinite lifetime &
radiation-dominated birth $\to$ $a_{0}=\mu^2/\sqrt{V_0}$ $\to$\\
& & radiation-like collapse; \textbf{finite total lifetime} \\[3pt]
$V_0=0$ & $a=Ae^{\alpha t}$, de Sitter &
radiation-dominated birth $\to$ $H\to\alpha$ de Sitter\\
& & (spontaneous radiation $\to$ dark-energy sequence) \\[3pt]
$V_0<0$ & $a=A/\sinh(\alpha t)$, big rip &
radiation-dominated birth $\to$ big rip as $V\to0^+$ \\
\bottomrule
\end{tabular}
\caption{Fate of the universe by the sign of $V_0$, with $\alpha=m/\sqrt2$.
For $C\neq0$ the early epoch always obeys the radiation law
\eqref{eq:radiation}; the crossover scale is $a_\times\sim\mu^2\sqrt C$.}
\label{tab:V0}
\end{table}

\begin{figure}[t]
\centering
\includegraphics[width=.8\textwidth]{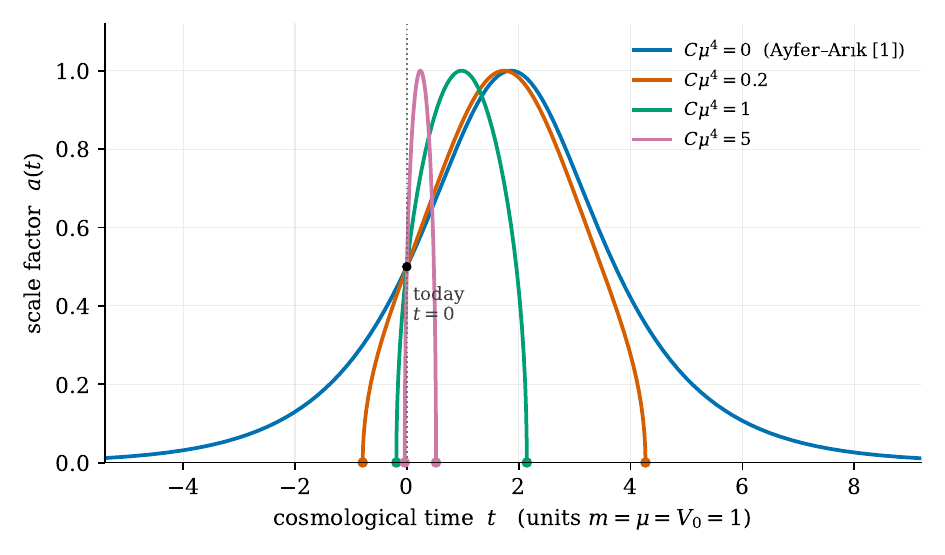}
\caption{Scale factor versus cosmological time ($m=\mu=V_0=1$). All curves
are normalized to the same $a(0)=a_0/2$ at $t=0$ (``today''), as in Fig.~1
of \cite{AyferArik2025}. $C\mu^4=0$: the eternal $1/\cosh$ solution of
\cite{AyferArik2025}. $C\neq0$: the universe is born at a finite time in the
past with the radiation law $a\propto\sqrt{t-t_b}$, reaches $a_{0}=1$ and
recollapses within a finite lifetime (dots: the $a\to0$ endpoints). The
larger the charge, the shorter the life.}
\label{fig:a}
\end{figure}

\begin{figure}[t]
\centering
\includegraphics[width=.8\textwidth]{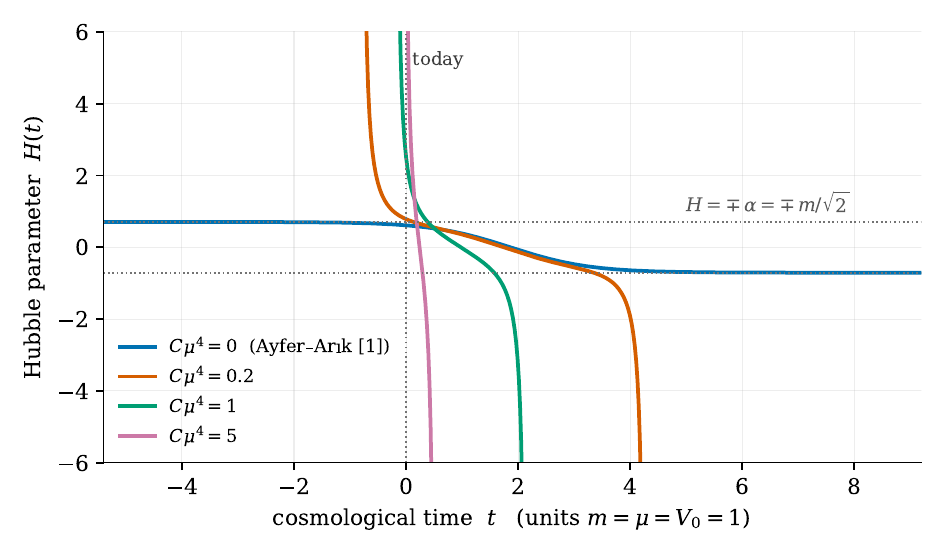}
\caption{Evolution of the Hubble parameter for the histories of
Fig.~\ref{fig:a}. The $C=0$ curve interpolates between the de Sitter
plateaux $\mp\alpha$; the $C\neq0$ curves reach radiation-like
singularities $H\to\pm\infty$ at the finite-time endpoints.}
\label{fig:H}
\end{figure}

\begin{figure}[t]
\centering
\includegraphics[width=.98\textwidth]{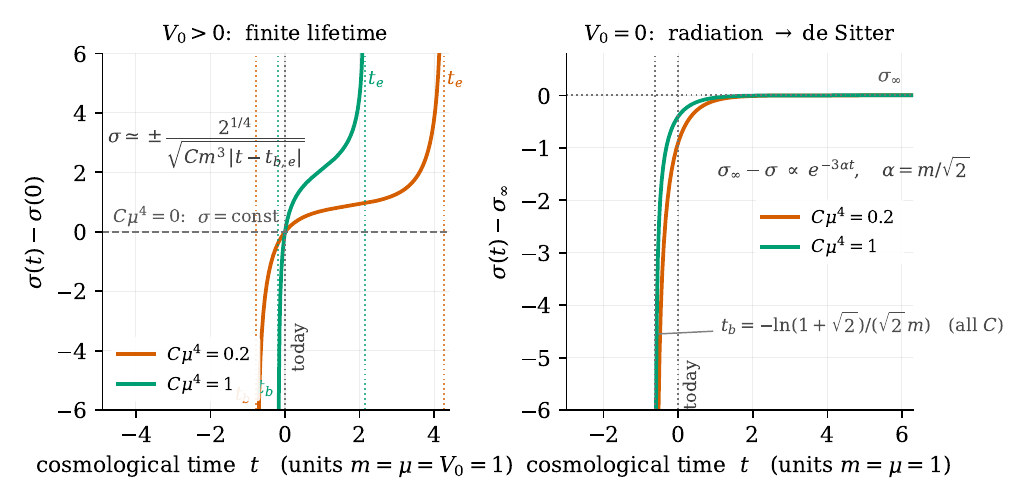}
\caption{Evolution of the canonical VEV field $\sigma=\rho/\sqrt2$
(physical vacuum expectation value $v=\rho=\sqrt2\,\sigma$) against
cosmological time; $t=0$ is ``today'' as in Fig.~\ref{fig:a}. Left, $V_0>0$:
between the endpoints $t_b$ and $t_e$ the field runs monotonically from
$-\infty$ to $+\infty$, diverging at the radiation-dominated birth and
collapse with the exact law
$\sigma\simeq\pm2^{1/4}\,(Cm^3\,|t-t_{b,e}|)^{-1/2}$ (verified numerically
to $8\times10^{-4}$); for $C=0$ the field does not evolve at all. Right,
$V_0=0$ (here $t=0$ is radiation/dark-energy equality): the field rises
from the big bang and \emph{freezes}, approaching its asymptote
$\sigma_\infty$ exponentially at the rate $3\alpha$ --- three $e$-folds of
the de Sitter expansion per $e$-fold of freezing --- so the electroweak
scale becomes constant in the dark-energy era. The birth time
\eqref{eq:age} is independent of the charge.}
\label{fig:vev}
\end{figure}

\begin{figure}[t]
\centering
\includegraphics[width=.98\textwidth]{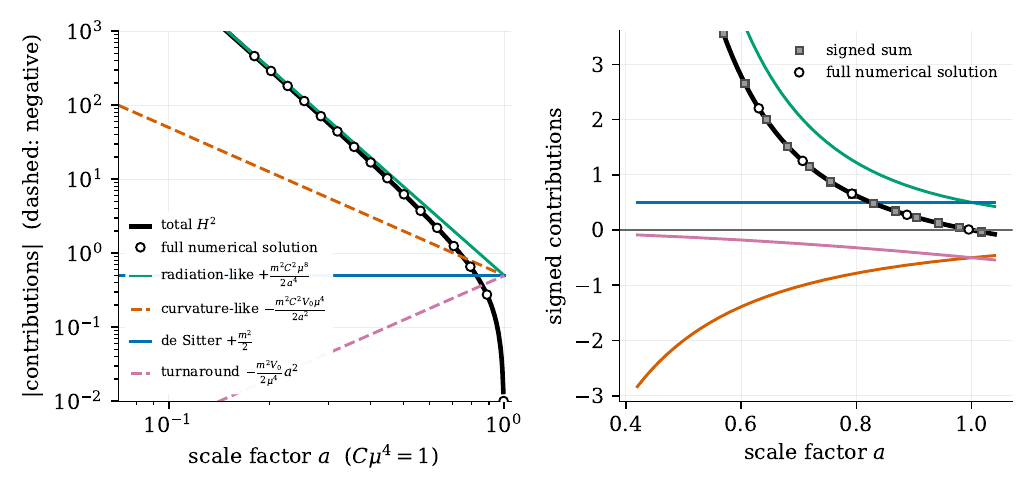}
\caption{\textbf{Comparison with the standard form of the Friedmann
equation.} The four contributions of \eqref{eq:friedmann}
($C\mu^4=1$, $m=\mu=V_0=1$, so that
$H^2=(1-a^2)(1+1/a^4)/2$). The $C=0$ part, $H^2=\alpha^2(1-a^2/a_0^2)$, was
found previously in \cite{AyferArik2025}. Left: magnitudes on a log--log
scale; dashed curves are absolute values of the \emph{negative}
(subtracted) contributions. Open circles: $H^2$ measured from the
integration of the full nonlinear system without using the formula
(relative difference $6.5\times10^{-7}$). Right: the same region on a
linear scale with \emph{signed} contributions; grey squares show the
algebraic sum of the four contributions, coinciding with the black total.}
\label{fig:H2}
\end{figure}

Figures \ref{fig:a}--\ref{fig:H} display the exact expansion histories, and
Figure~\ref{fig:vev} --- the central new plot of this paper --- the
evolution of the VEV-generating field itself, all against cosmological time
with $t=0$ chosen as ``today'' on the expanding branch, following the
normalization convention of Fig.~1 of \cite{AyferArik2025}. Figure~\ref{fig:H2}
compares the modified Friedmann equation with the standard form, overlaid
with a direct numerical integration of the full nonlinear system.

\section{Dynamical determination of the sign of the Higgs kinetic term}
\label{sec:sign}

The signature counting of Section~\ref{sec:model} fixes the signs of the
$\p$, $\phi^i$ and $\rho$ kinetic terms but leaves $s=f_{hh}=\pm1$ open. The
choice is invisible at the background level: for either sign the valley is
an exact solution and every formula of Section~\ref{sec:friedmann} holds
verbatim, only the canonical normalization of the valley field changing
($s=+1$: $\sigma=\rho/\sqrt2$; $s=-1$: $\sigma=\sqrt{3/2}\,\rho$). The two
theories separate at the level of the \emph{fluctuations} about the valley,
and only one of them possesses a stable broken vacuum. Throughout this
section we set $\varepsilon=+1$ in \eqref{eq:embedding} without loss of
generality.

\subsection{Orthogonal decomposition of the fluctuation}

In the $(h,\rho)$ block the field-space metric is
$G=\mathrm{diag}(s,-1)$. The valley curve, parametrized by $\bar\rho$, is
$(h,\rho)=(\bar\rho/\sqrt2,\,\bar\rho)$, with tangent vector
\begin{equation}\label{eq:tangent}
t^A=\frac{\dd}{\dd\bar\rho}\Bigl(\frac{\bar\rho}{\sqrt2},\,\bar\rho\Bigr)
=\Bigl(\frac1{\sqrt2},\,1\Bigr),
\qquad
f(t,t)=\frac s2-1 .
\end{equation}
We decompose an arbitrary fluctuation into its component along the valley
--- which merely reparametrizes the background --- and the component
$f$-orthogonal to it, which is the physical oscillation. Solving
$f(n,t)=0$, i.e.\ $s\,n^h/\sqrt2-n^\rho=0$, we may take
\begin{equation}\label{eq:normal}
n^A=\bigl(\sqrt2,\;s\bigr),
\qquad
k_\chi\equiv f(n,n)=2s-s^2=2s-1
=\begin{cases}+1, & s=+1,\\[2pt] -3, & s=-1,\end{cases}
\end{equation}
using $s^2=1$. The perturbed configuration is
\begin{equation}\label{eq:decomp}
\bigl(h,\rho\bigr)
=\Bigl(\frac{\bar\rho}{\sqrt2}+\sqrt2\,\chi,\;\;\bar\rho+s\,\chi\Bigr),
\end{equation}
with $\chi(t)$ the single physical fluctuation field; in terms of the naive
deviation $\delta\equiv h-\rho/\sqrt2$ one has, for $s=+1$,
$\delta=\chi/\sqrt2$.

\subsection{Curvature of the potential along the orthogonal direction}

Insert \eqref{eq:decomp} into $u\equiv h^2-\tfrac12\rho^2$:
\begin{align}
h^2&=\Bigl(\frac{\bar\rho}{\sqrt2}+\sqrt2\chi\Bigr)^2
=\frac{\bar\rho^{\,2}}2+2\,\bar\rho\,\chi+2\chi^2,\\
\tfrac12\rho^2&=\tfrac12\bigl(\bar\rho+s\chi\bigr)^2
=\frac{\bar\rho^{\,2}}2+s\,\bar\rho\,\chi+\tfrac12\chi^2,\\
u&=(2-s)\,\bar\rho\,\chi+\tfrac32\,\chi^2 .
\label{eq:uexpansion}
\end{align}
Since $V=\bar V+\lambda u^2$ and $u$ starts at first order, the quadratic
term of the potential in $\chi$ is $\lambda(2-s)^2\bar\rho^{\,2}\chi^2$,
i.e.
\begin{equation}\label{eq:Vchichi}
V_{\chi\chi}
=2\lambda\,(2-s)^2\,\bar\rho^{\,2}
=\begin{cases}
2\lambda\,\bar\rho^{\,2}, & s=+1,\\[2pt]
18\lambda\,\bar\rho^{\,2}, & s=-1 .
\end{cases}
\end{equation}

\subsection{First-order invariance and the exact linear equation}

A structural fact makes the linear analysis exact rather than approximate.
Insert \eqref{eq:decomp} into the kinetic form
$K=\tfrac12 s\,\dot h^2-\tfrac12\dot\rho^{\,2}$:
\begin{align}
s\,\dot h^2 &= s\Bigl(\frac{\dot{\bar\rho}}{\sqrt2}+\sqrt2\,\dot\chi\Bigr)^2
= s\,\frac{\dot{\bar\rho}^{\,2}}2+2s\,\dot{\bar\rho}\,\dot\chi
+2s\,\dot\chi^2,\\
\dot\rho^{\,2} &= \bigl(\dot{\bar\rho}+s\dot\chi\bigr)^2
=\dot{\bar\rho}^{\,2}+2s\,\dot{\bar\rho}\,\dot\chi+\dot\chi^2,\\
K &= \underbrace{\tfrac12\Bigl(\frac s2-1\Bigr)\dot{\bar\rho}^{\,2}}_{\bar K}
\;+\;\underbrace{\bigl(s\,\dot{\bar\rho}\,\dot\chi
- s\,\dot{\bar\rho}\,\dot\chi\bigr)}_{=\,0}
\;+\;\tfrac12\,k_\chi\,\dot\chi^2 .
\label{eq:Kexpansion}
\end{align}
The first-order term cancels identically --- this is the $f$-orthogonality
$f(n,t)=0$ at work. Likewise $V$ has no first-order term on the valley, by
\eqref{eq:uexpansion} with $u=0$ there. Consequently the potential, the
normalization condition, the scale factor $a=\mu^2/\sqrt V$, and the measure
$\sg=a^3$ are all \emph{unchanged at first order in} $\chi$: the
fluctuation does not react back on the background or on the emergent metric
at linear order (the metric, being algebraic, carries no independent
fluctuation of its own). Projecting the field equations
\eqref{eq:hred}--\eqref{eq:rhored} onto $n^A$ therefore yields a linear
equation whose coefficients are the \emph{exact} background quantities:
\begin{equation}\label{eq:linear}
\boxed{\;k_\chi\bigl(\ddot\chi+3H\dot\chi\bigr)
+V_{\chi\chi}(\bar\rho)\,\chi=0\;,}
\qquad\text{i.e.}\qquad
\ddot\chi+3H\dot\chi+\omega^2\chi=0,
\quad
\omega^2=\frac{V_{\chi\chi}}{k_\chi} ,
\end{equation}
the familiar damped-oscillator equation of scalar-field cosmology, exact at
linear order with no slow-roll or WKB assumption on the background. We have
verified symbolically that \eqref{eq:linear} coincides with the first-order
expansion of the $n^A$-projected full field equations for both signs
(Appendix~\ref{app:verification}). The construction is the analogue, for
the present constrained system, of the adiabatic/entropic decomposition of
multi-field inflation \cite{GordonWands2000}.

\subsection{The verdict: the Standard Model relation selects the sign}

The physics of \eqref{eq:linear} is decided by the sign of
$\omega^2=V_{\chi\chi}/k_\chi$. The same computation on the ridge $h=0$
gives the complementary entries; the full result is:
\begin{center}
\begin{tabular}{@{}lcc@{}}
\toprule
$\omega^2$ & valley $h^2=\rho^2/2$ & ridge $h=0$\\
\midrule
$f_{hh}=+1$ & $+2\lambda\rho^2$ \ (\textbf{stable}) &
$-2\lambda\rho^2$ \ (unstable)\\
$f_{hh}=-1$ & $-6\lambda\rho^2$ \ (unstable) &
$+2\lambda\rho^2$ \ (stable)\\
\bottomrule
\end{tabular}
\end{center}

For $s=+1$ the broken vacuum is stable, and the frequency of the Higgs
oscillation in cosmological time is
\begin{equation}\label{eq:higgsmass}
\boxed{\;\omega^{2}=2\lambda\,\bar\rho^{\,2}=2\lambda\,v^{2}
=m_H^{2}\;,}
\end{equation}
which is \emph{exactly} the Standard Model Higgs mass relation
\cite{ATLASCMS2012,PDG2024} --- a consistency condition the model satisfies
with no adjustable parameter. For $s=-1$ the broken vacuum decays at the
rate $\sqrt{6\lambda}\,v$; the stable configuration is then the symmetric
ridge $h=0$, electroweak symmetry never breaks, and the Higgs and fermion
masses disappear. The requirement that the broken phase be the physical
vacuum thus fixes the sign by itself:
\begin{equation}
f_{hh}=+1 ,
\end{equation}
i.e.\ the Standard Model fields belong to the positive-signature block of
$f_{ab}$, on the same side as the time-creating field.

\begin{figure}[t]
\centering
\includegraphics[width=.98\textwidth]{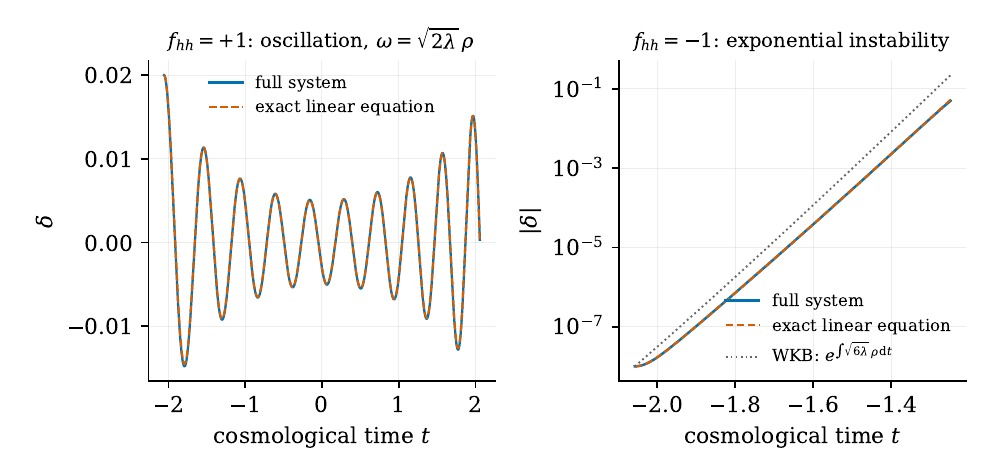}
\caption{Numerical verdict on the sign ($m=\mu=V_0=1$, $\lambda=4$,
$C\mu^4=0.1$, $\rho_0=4$). Left: for $f_{hh}=+1$ the deviation
$\delta=h-\rho/\sqrt2$ remains bounded and oscillatory; the full system and
the exact linear equation \eqref{eq:linear} coincide, with frequency
$\omega=\sqrt{2\lambda}\,\rho=m_H$. Right: for $f_{hh}=-1$ a $10^{-8}$
perturbation grows exponentially (log scale); dashed: exact linear
equation; dotted: WKB slope. The broken vacuum cannot survive this choice.}
\label{fig:sign}
\end{figure}

The numerical validation (Fig.~\ref{fig:sign}) is three-way. For $s=+1$
the frequency measured over 17 half-periods of the integrated full system
gives $\omega_{\rm meas}/\sqrt{2\lambda}\,\bar\rho=0.9996\pm0.0053$. For
$s=-1$ a seed of $10^{-8}$ grows by seven orders of magnitude within
$\Delta t\lesssim1$; the full system tracks the exact linear equation
\eqref{eq:linear} to $6\times10^{-5}$ in $\ln|\delta|$, and the measured
growth rate confirms the WKB value $\sqrt{6\lambda}\,\bar\rho$ at leading
order.

\section{The Higgs condensate and its backreaction}\label{sec:condensate}

With the sign fixed, the oscillation $\chi$ about the valley is the Higgs
particle of the model. Its cosmological evolution is completely governed by
the exact equation \eqref{eq:linear} and displays three phenomena:
freezing, an adiabatic amplitude law with a particle interpretation, and a
calculable backreaction on the expansion history.

\subsection{Freezing and the misalignment mechanism}

At early times ($a\ll a_\times$) the mass is small and the expansion fast:
$\omega=\sqrt{2\lambda}\,\bar\rho<H$. In this regime the friction term of
\eqref{eq:linear} dominates and $\chi$ is \emph{frozen} at its initial
value. Oscillation commences when $\omega\simeq H$
(Fig.~\ref{fig:backre}, left). This is precisely the misalignment mechanism
of axion and modulus cosmology \cite{PWW1983,Turner1983,Marsh2016}, here
realized by the Higgs fluctuation itself, with the novelty that the mass
grows because the VEV-generating field does.

\subsection{Adiabatic amplitude law and the particle interpretation}

In the oscillatory regime $\omega\gg H$, equation \eqref{eq:linear}
possesses the standard adiabatic invariant: writing
$\chi\simeq A(t)\cos\!\int\!\omega\,\dd t$, the comoving number of quanta
\begin{equation}
N\;\propto\;a^3\,\omega\,A^2=\text{const}
\end{equation}
is conserved. Solving for the amplitude with $a^3=\mu^6/\bar V^{3/2}$ and
$\omega=\sqrt{2\lambda}\,\bar\rho$,
\begin{equation}\label{eq:envelope}
A^2\;\propto\;\frac1{a^3\,\omega}
\;\propto\;\frac{\bar V^{3/2}}{\bar\rho}
\qquad\Longrightarrow\qquad
\boxed{\;A\;\propto\;\frac{\bar V^{3/4}}{\bar\rho^{\,1/2}}\;,}
\end{equation}
a parameter-free prediction which the numerical envelope follows with
$0.2\%$ scatter (Fig.~\ref{fig:backre}, left). The associated energy
density,
\begin{equation}\label{eq:rhochi}
\rho_\chi\equiv V_{\chi\chi}\langle\chi^2\rangle
=2\lambda\,\bar\rho^{\,2}\langle\chi^2\rangle
\;\propto\;\frac{m_H(t)}{a^3},
\end{equation}
is that of a gas of nonrelativistic particles with \emph{conserved comoving
number}, $n\propto a^{-3}$, and time-dependent mass
$m_H(t)=\sqrt{2\lambda}\,\bar\rho(t)$. Two limits are instructive: in the
radiation era $\bar\rho\propto\sqrt{\bar V}\propto a^{-1}$, so
$\rho_\chi\propto a^{-4}$ and the condensate redshifts like radiation; near
turnaround $\bar\rho\simeq$ const and $\rho_\chi\propto a^{-3}$,
matter-like. The condensate thus interpolates between radiation and matter
behaviour purely through the evolution of its mass.

\subsection{Backreaction: renormalization of the background constants}
\label{sec:backre}

At second order in the amplitude the condensate modifies the expansion
history. We measured this effect directly, with no perturbative
approximation, by integrating the full nonlinear $(h,\rho)$ system three
times --- no oscillation, amplitude $A$ ($\chi_0=0.05$), and amplitude
$A/2$ --- in units $m=\mu=V_0=1$ with $\lambda=2$, $C\mu^4=0.3$,
$\rho=25$ at the start of the window. All three histories were processed
with an \emph{identical} smoothing kernel in cosmological time, so that
averaging systematics are common mode and cancel in differences, and each
smoothed history was fitted to the background family
\begin{equation}
H^2=\frac{m_e^2}{2}\Bigl(1-V_{0e}\,a^2\Bigr)\Bigl(1+\frac{C_e^2}{a^4}\Bigr),
\end{equation}
the control run fixing the method residual at $\sim10^{-7}$. The measured
shifts of the effective constants are:
\begin{center}
\begin{tabular}{@{}lccc@{}}
\toprule
 & $\Delta m_e$ & $\Delta V_{0}$ & $\Delta C$\\
\midrule
amplitude $A$ ($\chi_0=0.05$) & $-1.02\times10^{-3}$ &
$+2.20\times10^{-3}$ & $+1.16\times10^{-3}$\\
$A^2$-scaling ratio (4 expected) & $4.01$ & $3.99$ & $3.99$\\
\bottomrule
\end{tabular}
\end{center}
The ratios in the second row compare the $A$ and $A/2$ runs: a genuine
second-order effect must scale by a factor of $4$, and all three shifts do
so to better than $1\%$. A search for a component \emph{outside} the family
(trial shapes $a^{-3}$, $a^{-4}$, $a^{-6}$, $\bar\rho/a^3$) returns
coefficients that fail the same $A^2$-scaling test (ratios $\approx0.85$):
at this precision the backreaction produces \textbf{no new power law} ---
it is entirely equivalent to the signed renormalization
$(m,V_0,C)\to(m_e,V_{0e},C_e)$ above.

Three remarks. (i) $\Delta C>0$: the condensate \emph{feeds} the
radiation-like component, exactly as expected from the $a^{-4}$ behaviour
of \eqref{eq:rhochi} in the radiation era --- an $a^{-4}$ energy input
appears, within the family, as a shift of the charge. (ii) $\Delta m_e<0$:
the asymptotic de Sitter rate $\alpha_e=m_e/\sqrt2$ \emph{decreases}.
(iii) The mixed signs show that the backreaction cannot be modelled as a
single-sign Einstein-like source, $\Delta H^2=\kappa\rho_\chi$ with a
constant $\kappa>0$: the emergent gravity couples to oscillation energy in
a structurally different way, as anticipated from the fact that here the
metric is sourced by the \emph{vanishing} of the total energy--momentum
tensor rather than by its value.

\begin{figure}[t]
\centering
\includegraphics[width=.98\textwidth]{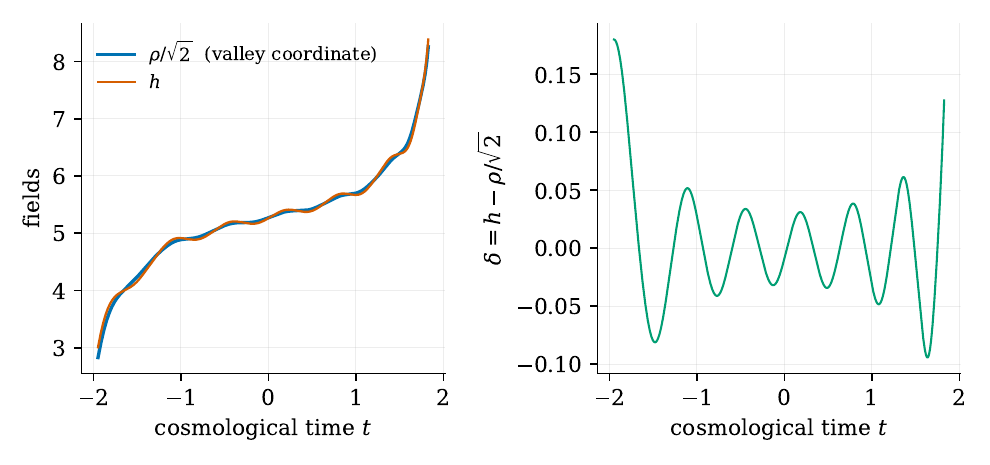}
\caption{Integration of the full $(h,\rho)$ system from valley-perturbed
initial data ($m=\mu=V_0=\lambda=1$, $C\mu^4=0.3$, $\delta=0.18$
initially; $t=0$ at the turnaround). Left: $h$ oscillates about the valley
coordinate $\rho/\sqrt2$. Right: the deviation $\delta$ remains bounded ---
the valley is stable; these oscillations are the Higgs condensate.}
\label{fig:higgs}
\end{figure}

\begin{figure}[t]
\centering
\includegraphics[width=.98\textwidth]{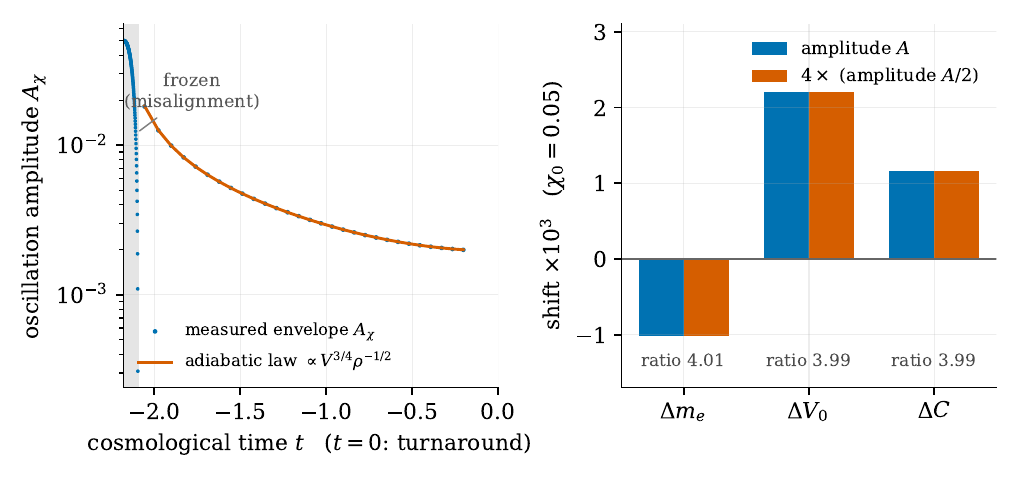}
\caption{Left: evolution of the deviation amplitude in cosmological time
--- the frozen (misalignment) epoch, oscillation onset at
$\omega\simeq H$, and the adiabatic law \eqref{eq:envelope} followed to two
parts in a thousand. Right: measured $O(A^2)$ shifts of the family
constants; the overlap of the $A$ and $4\times(A/2)$ bars demonstrates
second-order scaling.}
\label{fig:backre}
\end{figure}

\section{Verification}\label{sec:verification}

Every equation of this paper has been verified in two independent layers,
and we regard this as part of the result. The \emph{symbolic} layer
(\textsc{SymPy}) checks exact statements exactly: the metric equation
\eqref{eq:metric} component by component including the Higgs kinetic term;
the consistency identity of Section~\ref{sec:identity} both for arbitrary
field profiles and in the one-line form \eqref{eq:identitycheck}; the
exactness of the valley and ridge embeddings; the closed forms
\eqref{eq:sigmadot}--\eqref{eq:phidot}; the Friedmann equation
\eqref{eq:friedmann} in both the cosmological gauge and the static gauge of
Appendix~\ref{app:static}; and the decomposition data
\eqref{eq:normal}--\eqref{eq:Vchichi}, the first-order invariance
\eqref{eq:Kexpansion}, and the equality of \eqref{eq:linear} with the
linearized projected equations for both signs. The \emph{numerical} layer
(\textsc{SciPy}) integrates the full nonlinear system with no knowledge of
the closed forms and measures agreement: the valley is preserved to
$3.7\times10^{-12}$; $\dot\sigma$ matches the closed form to
$6.7\times10^{-10}$; the charge is conserved at machine precision; the
early-time expansion exponent is measured as $p=0.49989$ against the exact
$1/2$, with the coefficient of \eqref{eq:radiation} reproduced to four
digits; $H^2(a)$ from the simulation matches \eqref{eq:friedmann} to
$6.5\times10^{-7}$ (Fig.~\ref{fig:H2}); and the birth-to-crossover age
matches \eqref{eq:age}. The itemized list, with the precision of each
check, is given in Appendix~\ref{app:verification}.

\section{Conclusions and outlook}\label{sec:conclusions}

We have added a single ingredient --- a spacelike partner $\rho$ for the
Higgs-particle component, promoting the electroweak vacuum expectation
value to a dynamical field --- to the emergent-metric cosmology of Ayfer
and Ar{\i}k \cite{AyferArik2025}, and solved the resulting model
completely. The findings are as follows.

(1) The dynamics resides entirely in the $h$ and $\rho$ equations: by the
consistency identity of Section~\ref{sec:identity}, four of the six field
equations are automatic, leaving two physical degrees of freedom, the
background $\sigma$ and the Higgs oscillation $\chi$.

(2) The broken-symmetry valley $h^2=\rho^2/2$ is an exact, stable solution
of the full nonlinear system; on it the electroweak scale is carried by the
VEV-generating field, $v(t)=\rho(t)$, and the background is form-invariant
under the addition of the Higgs kinetic term.

(3) A single conserved charge solves the background in closed form and
generates, geometrically, the radiation-like $a^{-4}$ and curvature-like
$a^{-2}$ terms of the modified Friedmann equation \eqref{eq:friedmann} ---
components the pure Higgs cosmology lacked. For any $C\neq0$ the universe
acquires a radiation-dominated birth at finite cosmological time; for
$V_0=0$ the age of the universe at radiation/dark-energy equality is the
universal constant \eqref{eq:age}, and the expansion history realizes a
spontaneous radiation~$\to$~dark-energy sequence, of natural interest for
cosmic acceleration \cite{Riess1998,Perlmutter1999,PeeblesRatra2003} and
current discussions of evolving dark energy \cite{DESI2024,DESI2025} and
the Hubble tension
\cite{Planck2018,Riess2022,Verde2019,DiValentino2021,CosmoVerse2025}.

(4) The sign of the Higgs kinetic term is fixed dynamically: only
$f_{hh}=+1$ supports a stable broken vacuum, and that choice yields the
Standard Model Higgs mass relation $m_H^2=2\lambda v^2$
\cite{ATLASCMS2012,PDG2024} with no adjustable parameter.

(5) The Higgs condensate realizes a misalignment mechanism, obeys a
parameter-free adiabatic amplitude law interpreted as a conserved-number
gas of growing-mass Higgs particles, and backreacts on the expansion as a
signed renormalization of $(m,V_0,C)$ with $\Delta C>0$: the condensate
feeds the radiation component.

Several problems remain open. (i) Since $v(t)=\rho(t)$ evolves, the
constancy of dimensionless mass ratios must be established and confronted
with the bounds on varying constants \cite{Uzan2011}, in particular the BBN
and CMB constraints on a time-varying Higgs vacuum expectation value
\cite{YooScherrer2003}; in the $V_0=0$ branch the freezing of $\sigma$
(Fig.~\ref{fig:vev}, right) makes the late-time electroweak scale constant
automatically. (ii) A slow effective model separating the dynamical energy
transfer (the drift of $C_{\rm eff}$) from the initial displacement should
be constructed, and second-order averaged analytics should predict the
measured $(\Delta m_e,\Delta V_0,\Delta C)$. (iii) A matter-like $a^{-3}$
component is not detected in the present window; a dedicated
high-precision experiment focused on the turnaround region is required.
(iv) Observational bounds on the charge $C\mu^4$ --- for instance from the
scale of the pre-BBN radiation-like era --- deserve a separate study.
(v) At the quantum level, the connection with the emergent-gravity
amplitudes of \cite{CaroneErlichVaman2017,Erlich2018} suggests that the
long-distance limit of the present model contains a composite graviton;
whether the classical backreaction pattern found here survives
quantization is an open question.

\subsection*{Acknowledgements}
We thank T.~Ayfer and M.~Ergen for the foundational work on which this
paper builds, and \c{S}.~\c{S}ahin and T.~Tok for valuable discussions.

\appendix

\section{The static gauge and its equivalence}\label{app:static}

Instead of choosing the time coordinate to be cosmological time, one may
spend the fourth coordinate function by pinning the time-creating field to
a timelike coordinate $T$,
\begin{equation}\label{eq:staticgauge}
\p=\mu^2 T,\qquad \phi^i=\mu^2x^i ,
\end{equation}
the static gauge of the brane picture; $\p$ and $\phi^i$ then play exactly
the role of the St\"uckelberg fields of massive gravity in unitary gauge,
with $f_{ab}$ the reference metric \cite{dRGT2011}. Primes denoting
$\dd/\dd T$, the metric \eqref{eq:metric} becomes
\begin{equation}\label{eq:staticmetric}
g_{00}=\frac{\mu^4+h'^{\,2}-\rho'^{\,2}}{V}\equiv\frac{W^2}{V},
\qquad
g_{ij}=-\frac{\mu^4}{V}\delta_{ij},
\qquad
\sg=\frac{\mu^6W}{V^2},
\end{equation}
with $V=V_0+\tfrac12m^2\mu^4T^2$ on the valley. It must be emphasized that
$T$ is a coordinate label, not a clock: for $\rho'\neq0$ one has
$g_{00}\neq1$, and the physical cosmological time is recovered from
$\dd t=\sqrt{g_{00}}\,\dd T$. On the valley the conserved charge reads
$\sg\,g^{00}\dot\sigma_{(T)}=\mu^6\sigma'/(VW)=\mu^6C$, which is solved
algebraically:
\begin{equation}\label{eq:closedformT}
\sigma'(T)=\pm\frac{C\mu^2\,V}{\sqrt{1+C^2V^2}},
\qquad
W^2=\mu^4-\sigma'^{\,2}=\frac{\mu^4}{1+C^2V^2}>0 .
\end{equation}
Two properties are immediate: the emergent metric never degenerates
($W^2>0$ always), and as $|T|\to\infty$ the field saturates the
``speed limit'' $|\sigma'|<\mu^2$ and tracks the time-creating field,
$\sigma\simeq\pm\mu^2T=\pm\p$ --- compare the scalar speed limits of DBI
cosmology \cite{SilversteinTong2004}. Converting to cosmological time,
\begin{equation}
\frac{\dd\sigma}{\dd t}
=\frac{\sigma'}{\sqrt{g_{00}}}
=\sigma'\,\frac{\sqrt V}{W}
=C\,V^{3/2},
\qquad
a=\frac{\mu^2}{\sqrt V},
\qquad
H=-\frac{V'}{2\sqrt V\,W},
\end{equation}
and squaring $H$ with $V'=m^2\mu^4T$, $m^2\mu^4T^2=2(V-V_0)$ reproduces
\eqref{eq:H2V} exactly. All formulas of the main text follow; the two
gauges are equivalent descriptions of the same solution, as they must be
since $H$ and $a$ are observables. The practical advantage of the static
gauge is that \eqref{eq:closedformT} gives the background in closed form
without integrating any differential equation; the advantage of the
cosmological gauge of the main text is that no conversion between
coordinate labels and physical clocks is ever needed.

\section{Verification summary}\label{app:verification}

\paragraph{Symbolic layer (\textsc{SymPy} 1.14).}
Verified exactly, i.e.\ with vanishing residual after simplification: the
metric components \eqref{eq:metricgauge} and \eqref{eq:staticmetric} and
the trace relation $g^{\mu\nu}T_{\mu\nu}=2V$, with the componentwise
vanishing of $T_{\mu\nu}$ on the metric \eqref{eq:metric}, including the
Higgs kinetic term; the diffeomorphism identity
$\mu^2\,\delta\mathcal S/\delta\p+\dot h\,\delta\mathcal S/\delta h
+\dot\rho\,\delta\mathcal S/\delta\rho\equiv0$ in the static gauge for
\emph{arbitrary} $h$, $\rho$ profiles, and the one-line identity
\eqref{eq:identitycheck} in the cosmological gauge; the valley embedding
\eqref{eq:embedding} with \eqref{eq:closedformT} satisfying the $\p$, $h$
and $\rho$ equations simultaneously; the ridge equation \eqref{eq:ridge}
making the $\p$ equation vanish; the chain leading to
\eqref{eq:friedmann} in both gauges; the decomposition data
\eqref{eq:normal}--\eqref{eq:Vchichi}; the first-order invariance
\eqref{eq:Kexpansion}; and the equality of \eqref{eq:linear} with the
first-order expansion of the $n^A$-projected field equations, for both
signs $s=\pm1$.

\paragraph{Numerical layer (\textsc{SciPy}, \texttt{solve\_ivp}, tolerances
$10^{-12}$).}
Valley preservation $\max_t|h-\rho/\sqrt2|=3.7\times10^{-12}$ and
$\dot\sigma$ versus the closed form to $6.7\times10^{-10}$; charge
conservation at the $10^{-12}$ level; early-time log--log slope
$p=0.49989$ against the exact $1/2$, with the coefficient of
\eqref{eq:radiation} reproduced to $[0.9996,0.9999]$; $H^2$ from the full
simulation versus \eqref{eq:friedmann} to $1.9\times10^{-9}$ pointwise and
$6.5\times10^{-7}$ in the configuration of Fig.~\ref{fig:H2}; the field
evolution of Fig.~\ref{fig:vev} satisfying
$\dd\sigma/\dd t=CV^{3/2}$ and $a^3\dot\sigma=C\mu^6$ to $10^{-6}$, the
endpoint law of Fig.~\ref{fig:vev} to $8\times10^{-4}$, the freezing rate
$-3\alpha$ to $10^{-4}$, and the birth-to-crossover age
\eqref{eq:age} to $10^{-13}$ (two independent quadratures); sign analysis:
frequency ratio $0.9996\pm0.0053$ over 17 half-periods ($s=+1$),
full-versus-linear agreement $6\times10^{-5}$ in $\ln|\delta|$ and WKB
growth slope confirmed at leading order ($s=-1$); backreaction:
control-run family residual $\sim10^{-7}$, $A^2$-scaling ratios
$4.01/3.99/3.99$, envelope-law scatter $0.2\%$.

\section{The ridge solution}\label{app:ridge}

The embedding $h=0$, $\dot h=0$ satisfies \eqref{eq:hred} identically. On
this ridge $V=V_0+\tfrac12m^2(\p)^2+\tfrac\lambda4\rho^4$; the $\rho$
equation does not integrate to a charge, because
$\partial V/\partial\rho\big|_{h=0}=\lambda\rho^3\neq0$, and in the static
gauge of Appendix~\ref{app:static} it reads
\begin{equation}\label{eq:ridge}
\rho''
=\frac{\bigl(\mu^4-\rho'^{\,2}\bigr)
\bigl(\lambda\rho^3+m^2 T\,\rho'\bigr)}{V}\;,
\end{equation}
the $\p$ equation again being automatic (verified symbolically). Being a
maximum of $V$ in the $h$ direction, the ridge is unstable for the physical
sign $f_{hh}=+1$ (Section~\ref{sec:sign}), and the physical evolution is
that of Sections~\ref{sec:valley}--\ref{sec:friedmann}.

\end{document}